\documentstyle[12pt]{article}
\def\be{\begin{equation}}
\def\ee{\end{equation}}
\def\bq{\begin{eqnarray}}
\def\eq{\end{eqnarray}}

\begin{document}

\begin{titlepage}
\vspace{0.3in}
\begin{flushright}
CEBAF-93-03//
January 1993
\end{flushright}
\vspace{0.4in}
\begin{center}
{\Large \bf Soft Modes Contribution into Path Integral.}
\vspace{0.2in}
{\bf V.M.Belyaev   \\ }

Institute of Theoretical and Experimental Physics
                         Moscow, 117259 \\
and\\
Continuous Electron Beam Accelerator Facility\\
12000 Jafferson Ave, Newport News, Virginia 23606, USA\\
\vspace{0.2in}
{\bf   Abstract  \\ }
\end{center}
A method for nonperturbative  path integral calculation
is proposed.
 Quantum mechanics as a
simplest example of a quantum field theory is considered.
All modes are decomposed  into hard (with frequencies $\omega^2 >
\omega^2_0$) and soft (with frequencies $\omega^2 <
\omega^2_0$) ones, $\omega_0$ is a some parameter.
 Hard modes contribution is considered
by  weak coupling expansion.  A low energy effective
Lagrangian for soft modes is used.
 In the  case of soft modes
we apply a strong coupling expansion.
To realize this expansion
a  special basis in functional space of trajectories
  is  considered.
 A good convergency of proposed procedure
in the  case of  potential $V(x)=\lambda x^4$
is demonstrated.
Ground  state energy of the unharmonic  oscillator is calculated.

\end{titlepage}

Path integral formalism \cite{fei} is one of the most useful
tools to study a quantum field
theory.
However there is a serious problem to go out of boundaries of
a perturbative theory.
There are  instanton calculations \cite{inst},
  a   lattice calculation method \cite{lat} and
 variational
approach which can be used in the case of quantum field theory \cite{var}
and sometimes it  is possible  to find nonperturbative exact results
using symmetries of a  quantum field model \cite{susy}.

 Here
we propose an alternative method for nonperturbative
 path integral computations.
 All modes are decomposed
 into hard (with $\omega^2 > \omega_0^2$) and soft
(with $\omega^2 < \omega^2_0$) modes where $\omega_0$ is a some parameter.
 It is clear that when a frequency
is enough large then we can consider a potential term as a perturbation
and  use a conventional pertubative theory.
Thus we can find  an effective Lagrangian \cite{eff}
for soft modes
using  wellknown  perturbative theory.
To find a
 calculation procedure for soft modes  we assume that
 the frequencies of these modes are enough small and in the
leading approximation we
can neglect a kinetic term and all other terms with derivatives in the
effective lagrangian and use  a strong coupling expansion.
To realize a strong coupling expansion a special basis
for  trajectories in
functional space is  suggested  and in this basis a regular
scheme for the soft modes contribution is formulated in the
Section 3.

Here we consider quantum
mechanics as a simplest example of a quantum field theory.
It is possible, that this method can be applied in a
quantum field theory but it requires
additional investigations, particularly,
to take into consideration a
renormalization and  a gauge
invariance
(in the case of a gauge theory).
 Also a problem of a convergency of this
procedure is opened and we just demonstrate a respectively
 good convergency
in the case of unharmonic oscillator with a potential
$V(x)=\lambda x^4$.

In the next Section, a path integral and a basis in
a functional space of trajectories are considered.
In the Section 2,  we formulate a procedure of
nonperturbative calculation of soft modes contribution in the limit
of large coupling constant. The soft modes contribution is calculated
in the case of quantum mechanics with a potential $\lambda x^4$.
Then  in  Section 3  we find the first  correction due to the
kinetic term.
Ground  state energy  of  the system is calculated in  Section 4.
Here  we  take into account  2-loop  effective potential.
In Conclusions  we discuss uncertanties  of the  calculations
and  possibility  to use  the procedure  in other  field theories.

In this paper we consider quantum mechanics in euclidean formalism.
\newpage
\section{Path Integral}

We consider the following path integral \cite{fei}
\bq
<x_f\mid e^{-\hat{H}t_0}\mid x_i>={\cal N}^{-1}\int {\cal D}x(t)
e^{-\int_0^{t_0}{\cal L}(x(t))dt}
\label{1}
\eq
where ${\cal L}(x(t))=\frac{1}{2}(\frac{dx}{dt})^2+V(x)$, $x(0)=x_i$,
$x(t_0)=x_f$, $\hat{H}$ is a hamiltonian of a system, ${\cal N}$ is a
normalization factor.

Here we are interesting in  a lowest state energy  $\varepsilon_0$
 and it is convenient to consider the limit
$t_0 \rightarrow \infty$ and to find a trace over $x$ in (\ref{1}), i.e.
$x_i(0)=x_f(t_0)$,
\bq
{\cal Z}=\int dx <x\mid e^{-\hat{H}t_0}\mid x>=\int dx <x\mid n>
e^{\varepsilon_nt_0}<n\mid x>_{\mid t \rightarrow \infty}
\label{2}
\eq
\bq
=\int dx \mid\Psi_0(x)\mid^2e^{-\varepsilon_0t_0}=e^{-\varepsilon_0t_0}
\nonumber
\eq
where $\varepsilon_n$ is an energy of $n-$th state, and $\varepsilon_0$ is
the lowest energy of the system.
The factor $\cal N$ is
$\int {\cal D}x(t)e^{-\int_0^{t_0}\frac{1}{2}(\frac{dx}{dt})^2dt}$.

In a perturbative theory the following basis for trajectories  is  used
\bq
x(t)=\sum_{n=-\infty}^{+\infty}C_ne_n(t)
\label{3}
\eq
where $e_n(t)=\frac{1}{\sqrt{t_0}}e^{i\omega_nt}$,
$\omega_n=\frac{2\pi}{t_0}n$,
$C_n=C_{-n}^*$.

This basis $\{ e_n\}$  has the  following normalization
\bq
<e_n\mid e_m> = <e_n^*e_m>=\int_0^{t_0}e^*_n(t)e_m(t)dt=\delta_{mn}
\label{3a}
\eq
Therefore PI (\ref{1}) in basis (\ref{3}) has the following form
\bq
{\cal Z}={\cal N}^{-1}\int \prod_{n=-\infty}^{+\infty}\frac{dC_n}{\sqrt{2\pi}}
e^{-<{\cal L}(\sum_nC_ne_n)>}
\label{4}
\eq
Here we use the denotation: $<f(t)>=\int_0^{t_0}f(t)dt$.

Hard modes are taken into consideration by conventional perturbative
theory and  after  integration over hard modes we obtain a low energy
effective Lagrangian for the soft ones.
Soft modes are considered in context of a strong coupling expansion
and  a soft modes kinetic term  is considered as a perturbation as well
as all  other terms with derivatives in the effective Lagrangian.
However a computation of this  contribution
is rather difficult even if we neglect the kinetic term.
It is known the way to use a strong coupling expansion in a lattice
theory
where  we should to choose coupling constants and parameters of
a lattice to have a correct continuum limit.
Here we propose an alternative approach for strong coupling expansion.
We do not change a theory but only change a  basis in functional
space of trajectories:
\bq
x(t)=\sum_{\mid n \mid < N}B_nE_n(t)+\sum_{\mid n\mid >N}C_ne_n(t)
\label{5}
\eq
\bq
\omega_0=\frac{2\pi}{t_0}N;\;\;
\nonumber
\eq
\bq
<E_{m_1}E_{m_2}...E_{m_n}>=(\omega_0/\pi)
^{(n-2)/2}A_n\delta_{m_1m_2}\delta_{m_1m_3}...\delta_{m_1m_n}
\nonumber
\eq
where $A_n$ is a some number which depends
on a
choice of the
basis $\{E_n\}$, $n>1$. (Notice, that two subspaces
$\{ e_n\}$ $|n|>N$
and ${E_n}$ $|n|<N$ are not orthogonal to each other.)

The most  important feature of the subspace $\{ E_n\}$ is the
fact that in this subspace there is a factorization of the path integral
  if we neglect terms with derivatives in the action.
It gives us a possibility  to apply a strong coupling expansion.
Soft modes belong to the subspace $\{ E_n\}$  only.
But  there are  hard  modes  in this subspace too.
In the next Section, a regular procedure  for calculation of  pure  soft modes
contribution is formulated.
Notice, that this basis $E_n$ breaks
translational invariance of the path integral. This invariance
is restored when we subtract hard modes contribution out of
 the subspace $\{ E_n\}$.

Here we use one of the possible choices for the basis $\{ E_n\}$
\bq
E_n(t)=\frac{1}{\sqrt{\Delta t}}\Theta (t-t_0/2-n\Delta t)\Theta
(t_0/2+(n+1)\Delta t -t)
\label{6}
\eq
where $\Delta t=\pi/\omega_0$. It is obviously that in this basis we have
$A_n=1$.

Below we use the following denotations:\\
greek letters: $\mu$, $\nu$,..$=0,\pm 1,..,\pm N$;\\
small letters: $m$, $n$,..$=\pm(N+1),\pm(N+2),..$;\\
large letters: $M$, $L$,..$=0,\pm 1,..\;\infty$\\

Let us show that
\bq
{\cal Z}=\int \prod_n\frac{dC_n}{\sqrt{2\pi}}\prod_\mu
\frac{dB_\mu}{\sqrt{2\pi}}
e^{-<{\cal L}(\sum (C_ne_n+B_\mu E_\mu ))>}\mid J \mid
\label{8}
\eq
where
\bq
J=det ( <e_\mu E_\nu >)
\label{9}
\eq
Using that $E_\mu=<E_\mu e_M>e_M$   we have from (8)
\bq
{\cal Z}=\int \prod_n\frac{dC_n}{\sqrt{2\pi}}\prod_\mu
\frac{dB_\mu}{\sqrt{2\pi}}
e^{-<{\cal L}(\sum_nC_ne_n+\sum_\mu B_\mu
\sum_M<E_\mu e^*_M>e_M)>}\mid J \mid
\label{10}
\eq
Then shifting $C_n$ we cancel terms with $e_n$ in the sum over $M$ and $\mu$
and
 obtain
\bq
{\cal Z}=\int \prod_n\frac{dC_n}{\sqrt{2\pi}}\prod_\mu
\frac{dB_\mu}{\sqrt{2\pi}}
e^{-<
{\cal L}(\sum C_ne_n+\sum_\mu\sum_\nu B_\mu <E_\mu e^*_\nu >)>}
\mid J \mid
\label{11}
\eq
Using the following variables
\bq
C_\mu = \sum_\nu B_\nu <E_\nu e^*_\mu>
\eq
and taking into account the Jakobian  we reproduce
eq.(\ref{4}).

Let us calculate $|J|$. The simplest way is to consider the
determinant of
the  following matrix
\be
M_{\mu\nu}=<e^*_\mu E_\rho ><E_\rho e_\nu >
\label{d1}
\ee
\bq
|J|=\sqrt{\det (<e^*_\mu E_\rho ><E_\rho e_\nu >)}
\eq
Where $M_{\mu\nu}$ is
\be
M_{\mu\nu}=\frac{1}{\Delta t t_0}\sum_\rho
\int_{\rho\Delta t+t_0/2}^{(\rho +1)\Delta t+t_0/2}e^{-i\omega_\mu t_1}dt_1
\int_{\rho\Delta t+t_0/2}^{(\rho +1)\Delta t+t_0/2}e^{+i\omega_\nu t_2}dt_2=
\label{d2}
\ee
\bq
=\frac{(e^{-i\omega_\mu\Delta t}-1)(e^{+i\omega_\nu\Delta t}-1)}{\Delta t
t_0 \omega_\mu\omega_\nu}\sum_{\rho}e^{-i(\omega_\mu -\omega_\nu)\Delta t
\rho}
\eq
and the determinant has the following form
\be
\det (M_{\mu\nu})=\prod_{\mu ,\nu}
\left(
\frac{(e^{-i\omega_\mu\Delta t}-1)(e^{+i\omega_\nu\Delta t}-1)}
{\Delta t^2
 \omega_\mu\omega_\nu}\right) \det (N_{\mu\nu})
\label{d3}
\ee
\bq
N_{\mu\nu}=\frac{\Delta t}{t_0}\sum_\rho e^{-i(\omega_\mu
-\omega_\nu )\Delta t}
\eq

There are two different cases for matrix elements $N_{\mu\nu}$:
diagonal ($\mu =\nu$) and nondiagonal ($\mu\neq\nu$). When $\mu =\nu$
then we have
\be
N_{\mu\nu\; |\mu =\nu}=\frac{\Delta t}{t_0}
\sum_\rho 1=1
\label{d4}
\ee
Nondiagonal elements are
\be
N_{\mu\nu\; |\mu \neq\nu}=\frac{\Delta t}{t_0}
\sum_\rho e^{-i(\omega_\mu -\omega_\nu)\Delta t\rho}=
0
\label{d5}
\ee
 Here we use the periodical boundary condition for $\{ e_\mu \}$

Thus from  (\ref{d4}) and (\ref{d5}) we obtain that
\be
\det (N_{\mu\nu})=1
\label{d7}
\ee
and
\be
|J|=\exp \left(\frac{1}{2}
(\sum_{\mu ,\nu}\ln (\frac{(e^{-i\omega_\mu\Delta t}-1)
(e^{i\omega_\nu\Delta t}-1)}{\Delta t^2\omega_\mu\omega_\nu}))\right)
\label{d8a}
\ee
\bq
=\exp \left(\frac{1}{2}\sum_\mu \ln (\frac{2((1-\cos (\omega_\mu\Delta t))}
{(\omega_\mu\Delta t)^2}\right)
\nonumber
\eq
\bq
=\exp (-\frac{\omega_0t_0}{2\pi}j)
\nonumber
\eq
where
\be
j=-\int_0^\pi\frac{dx}{\pi}\ln (\frac{2(1-\cos (x))}{x^2})=2(\ln (\pi )-1)=
0.289...
\label{d9}
\ee
\newpage
\section{Soft Modes Contribution}

Let us start to study  a
quantum mechanics with a potential
$V(x)=\lambda x^4$.
The Lagrangian has a form
\be
{\cal L}=\frac{1}{2}(\frac{dx}{dt})^2+\lambda x^4
\label{b1}
\ee
In terms of our basis $\{ E_\mu\}+\{ e_n\}$ the path integral is
\bq
{\cal Z}=\frac{1}{N}\int \prod_n\frac{dC_n}{\sqrt{2\pi}}\prod_\mu
\frac{dB_\mu}{\sqrt{2\pi}}|J|
\label{b2}
\eq
\bq
\times
\exp \{ -[\frac{1}{2}|C_n|^2\omega_n^2+C_n\omega_n^2<e_n E_\mu >B_\mu+
\frac{1}{2}B_\mu <E_\mu e_N>\omega_N^2<e_N^*
 E_\nu>B_\nu
\nonumber
\eq
\bq
+\lambda (B_\mu^4<E_\mu^4>+4B_\mu^3<E_\mu^3 e_n>C_n+
6B_\mu^2<E_\mu^2 e_me_n>C_mC_n
\nonumber
\eq
\bq
+4B_\mu <E_\mu e_me_ne_k>C_mC_nC_k+<e_me_ne_ke_l>C_mC_nC_kC_l)]\}
\nonumber
\eq
To cancel linear terms for hard modes ($C_n$) in the kinetic term
 we make a shift:
$C_n=C_n-<e_n^*E_\mu>B_\mu$. This shift we have made in (\ref{8}).
After the shift we have
\be
{\cal Z}=\frac{1}{N}\int \prod_n\frac{dC_n}{\sqrt{2\pi}}\prod_\mu
\frac{dB_\mu}{\sqrt{2\pi}}|J|
\label{b3}
\ee
\bq
\times
\exp \{ -[\frac{1}{2}|C_n|^2\omega_n^2+
\frac{1}{2}B_\mu <E_\mu e_\rho>\omega_\rho^2<e_\rho^* E_\nu>B_\nu+
\lambda (B_\mu^4<E_\mu^4>
\nonumber
\eq
\bq
-4B_\mu^3<E_\mu^3 e_\rho ><e_\rho^* E_\nu>B_\nu +
6B_\mu^2<E_\mu^2 e_me_n><e_m^*E_\nu >B_\nu <e_n^*E_\rho >B_\rho
\nonumber
\eq
\bq
-4B_\mu<E_\mu e_me_ne_k><e_m^*E_\nu >B_\nu <e_n^*E_\rho >
B_\rho <e_k^*E_\lambda >
B_\lambda
\nonumber
\eq
\bq
+<e_me_ne_ke_l><e_m^*E_\mu >B_\mu <e_n^*
E_\nu >B_\nu <e_k^*E_\rho >
B_\rho <e_l^*E_\lambda >B_\lambda
\nonumber
\eq
\bq
+(terms\; with\;C_n)]\}
\nonumber
\eq

In this Section we do not consider the hard modes
 and neglect kinetic term for
soft modes. To calculate the contribution
we expand (\ref{b3}) over a number of
projections from subspace $\{ E_\mu\}$ into subspace $\{ e_n\}$ and back.
This procedure  corresponds to a
regular  subtraction of hard modes
out of subspace $\{E_n\}$.
These
projections decrease a norm of the vector $x(t)$ in a factor
$\kappa <1$ which depends on the vector in functional space.
When we integrate over
all subspace $\{ E_n\} $ we can expect that an
effective value of this factor is
enough small. To estimate $\kappa$ we can
consider the jakobian $|J|$ which is equal
to unit in the  case of ortohonality between $\{E_n\}$ and
$\{e_\nu\}$ subspaces.
  A deviation $|J|$ from $1$ is a measure
of nonortohonality between these two subspaces.
In our case $|J|=\exp (-\frac{\omega_0t_0}{2\pi}j)$
which can be absorbed by rescaling $B_\mu\rightarrow B_\mu e^{-j/2}$.
It is reasonable to suppose that $\kappa = j/2 = 0.145...$.

Thus, in the leading order of our expansion for soft modes  we
have
\be
{\cal Z}_{soft}=e^{-\varepsilon_s^{(0)}t_0}=
\frac{1}{{\cal N}_{soft}}e^{-\frac{\omega_0t_0}{2\pi}j_0}
(\int_{-\infty}^{+\infty}\frac{dB}{\sqrt{2\pi}}
e^{-\lambda B^4/\Delta t})^{\frac{\omega_0t_0}{\pi}}
\label{b4}
\ee
where
\bq
{\cal N}_{soft}=\int \prod_\mu
\frac{dC_\mu}{\sqrt{2\pi}}e^{-\frac{1}{2}
|C_\mu |^2\omega_\mu^2}
\nonumber
\eq
is the normalization factor for soft modes.
In (\ref{b4}) we neglect nonortohonality between
$\{ E_\mu\}$ and $\{ e_n\}$ subspaces. In this case
$|J|=1$ and $j=j_0=0$.

To take into account the
first corrections of the
the expansion over numbers of the projections for the jakobian
$|J|$  it is useful to represent $j$ in the following
form
\bq
j=-\frac{\pi}{\omega_0t_0}tr\ln (<E_\mu e_\rho^* >
<e_\rho E_\nu >)
\nonumber
\eq
\be
=-\frac{\pi}{\omega_0t_0}tr\ln (<E_\mu e^*_N><e_NE_\nu >-
<E_\mu e_n><e_nE_\nu >)
\label{dd1}
\ee
\bq
=-\frac{\pi}{\omega_0t_0}tr\ln (\delta_{\mu\nu}-
<E_\mu e_n^*><e_nE_\nu >)
\nonumber
\eq
\bq
=\frac{\pi}{\omega_0t_0}tr (0+<E_\mu e_n^*><e_nE_\nu >
\nonumber
\eq
\bq
+\frac{1}{2}<E_\mu e_n^*><e_nE_\rho ><E_\rho e_m^*><e_mE_\nu >+...\; )
\nonumber
\eq
\bq
=j_0+j_1+j_2+...
\nonumber
\eq
Here
\bq
j_0=0
\nonumber
\eq
\bq
j_1=\frac{2}{\pi}\int_\pi^\infty\frac{1-\cos (x)}{x^2}dx=0.227...
\nonumber
\eq

Thus we have
\bq
{\cal Z}_{soft}=e^{-\varepsilon_s^{(0)}t_0}
\label{b5}
\eq
\bq
\varepsilon_s^{(0)}=\frac{\omega_0}{\pi}
(j_0/2 - \ln (\frac{\Gamma (1/4)}{2(4\pi \omega_0\lambda)^{1/4}})-
\ln (\omega_0)+1)
\label{b6}
\eq
\bq
=\frac{3\omega_0}{4\pi}\left( 1-\ln (\frac{\omega_0\Gamma (1/4)^{4/3}}
{4(\pi e\lambda )^{1/3}})\right)
\nonumber
\eq

Let us find the first correction for $\varepsilon_s$ in the expansion over
number of projections from subspace $\{ E_\mu\}$ into $\{ e_n\}$ and back. It
is clear that the first correction appears due to the term
$-4\lambda B_\mu^3<E_\mu^3e_n><e_nE_\nu>B_\nu$ in the action for soft modes
(\ref{b3}). This term gives the following contribution into the path
integral
\bq
{\cal Z}^{(1a)}_s=\exp (-(\varepsilon^{(0)}_s+\varepsilon^{(1a)}_s)t_0)=
{\cal Z}^{(0)}_s(1-\varepsilon^{(1a)}t_0)=
\label{b7}
\eq
\bq
\frac{1}{{\cal N}_s}\int \prod_\mu\frac{dB_\mu}{\sqrt{2\pi}}|J|
\exp (-\lambda B_\mu^4<E_\mu^4>)(1+4\lambda B_\mu^3<E_\mu^3e_n><e_nE_\nu>B_\nu
)
\nonumber
\eq
here we use the denotation $\varepsilon_s^{(1a)}$  because
there is another correction of the same order of the expansion.

{}From (\ref{b7}) we
obtain
\be
\varepsilon_s^{(1a)}=-\frac{1}{t_0} \times
\label{b8}
\ee
\bq
\times\sum_\mu\left(\int\frac{dB_\mu}{\sqrt{2\pi}}
(4\lambda B_\mu^3<E_\mu^3e_n><e_n^*E_\nu>B_\nu)\right)
\left(
\int \prod_\mu\frac{dB_\mu}{\sqrt{2\pi}}\right)^{-1}
\nonumber
\eq
\bq
=-4\frac{\omega_0}{\pi}\frac{\Gamma (5/4)}{\Gamma (1/4)} j_1
\nonumber
\eq
where $j_1$ is determined in (\ref{b5}).

Let us consider a contribution of the following term in
the action (\ref{b3}):
\be
6\lambda B_\mu^2<E_\mu^2e_me_n><e_n^*E_\nu >B_\nu <e_m^*E_\rho >B_\rho
\label{b11}
\ee
This term gives the following correction for
$\varepsilon_s$
\be
\varepsilon_s^{(1b)}=\frac{\omega_0}{\pi}
6\lambda j_1\frac{\omega_0}{\pi}(\int\frac{dB}{\sqrt{2\pi}}B^2e^{-\lambda
B^4\frac{\omega_0}{\pi}})^2
(\int\frac{dB}{\sqrt{2\pi}}e^{-\lambda B^4\frac{\omega_0}{\pi}})^{-2}=
\label{b12}
\ee
\bq
\frac{\omega_0}{\pi}6\left( \frac{\Gamma (3/4)}{\Gamma (1/4)}
\right) ^2j_1=
\frac{\omega_0}{\pi}(0.685...)j_1
\nonumber
\eq
Thus we have
\be
\varepsilon_s\simeq
\varepsilon_s^{(0)}+\varepsilon_s^{(1a)}+\varepsilon_s^{(1b)}
\label{b14}
\ee
\bq
=\frac{3\omega_0}{4\pi}\left( 1-\ln (\frac{\omega_0\Gamma (1/4)^{4/3}}
{4(\pi e\lambda)^{1/3}e^{0.056}})\right)
\nonumber
\eq
where the first corrections of our expansion ($j_1$, $\varepsilon_s^{(1a)}$
and $\varepsilon_s^{(1b)}$) are taking into account in a
factor $e^{0.056}$. All other terms of the action (\ref{b3}) correspond to
the higher corrections of the  expansion.

Thus,
the next to the leading order of the expansion
gives a  small contribution into the
$\varepsilon_s$ ($\sim 6\%$) and we can expect that next
corrections
of the expansion are small.

The maximal value for $\varepsilon_s$ is
\be
\varepsilon_s=
\frac{3\omega_0}{4\pi}=0.35\lambda^{1/3};\;at\;\;\omega_0=
\omega^*= e^{0.056}\left(
\frac{64\pi e \lambda}
{\Gamma^4 (1/4)}\right)^{1/3}\simeq 1.55\lambda^{1/3}
\label{b15}
\ee

The dependence of $\varepsilon_s(\omega_0)$ on $\omega_0$ is depicted in
Fig.1 where we put $\lambda =1$.
The exact value for ground state energy is $0.66..\lambda^{1/3}$.
which is about two times larger than  the maximal value for $\varepsilon_s$.

\section{Soft Modes Kinetic Term Contribution}

Let us take into account a leading contribution of a kinetic term for
the soft modes into $\varepsilon_s$.
Then this correction  for the
 energy $\varepsilon_s$ is
\be
\varepsilon_s^{k1}=
\frac{\omega_0}{\pi}\left( \frac{1}{2}\int dBB^2<E_1e_\rho >
\omega_\rho^2<e_\rho^*E_1>e^{-\lambda B^4\frac{\omega_0}{\pi}}\right)
\label{k1}
\ee
\bq
\times\left( \int dBe^{-\lambda B^4\frac{\omega_0}{\pi}}\right) ^{-1}
\nonumber
\eq
where
\be
<E_1e_\rho >\omega_\rho^2<e_\rho^*E_1>=
\sum_\rho\frac{\omega_\rho^2}{t_0\Delta t}
\int_0^{\Delta t}e^{i\omega_\rho t_1}dt_1
\int_0^{\Delta t}e^{-i\omega_\rho t_2}dt_2
\label{k2}
\ee
\bq
=\sum_\rho\frac{\omega_0}{\pi}\frac{\omega_\rho^2}{t_0}
\frac{2(1-\cos (\omega_\rho\Delta t))}{\omega_\rho^2}
\nonumber
\eq
\bq
=\frac{\omega_0}{\pi}\int_0^{\omega_0}\frac{d\omega}{\pi}
2(1-\cos (\frac{\omega}
{\omega_0}\pi))=2\left(\frac{\omega_0}{\pi}\right)^2
\nonumber
\eq

Then from (\ref{k1}) and (\ref{k2}) we have
\be
\varepsilon_s^{k1}=
\frac{\omega_0}{\pi}\left(\frac{1}{2}\frac{2\omega_0^2}{\pi^2}
\sqrt{\frac{\pi}{\omega_0\lambda}}\frac{\Gamma (3/4)}{\Gamma(1/4)}\right)
\label{k3}
\ee
\bq
=\frac{\omega_0}{\pi}\left(\frac{\Gamma (3/4)}{\Gamma (1/4)}
\sqrt{\frac{\omega_0^3}{\pi^3\lambda}}\right)
\nonumber
\eq

At $\omega_0=\omega^*$
we have
\be
\varepsilon_s^{k1}\simeq 0.14\varepsilon_s
\label{k4}
\ee
The next correction  at
$\omega_0=\omega^*$ is about $1\%$ and
we do not take it into consideration.

\section{The Ground State Energy}

To have  a  reliable result  for  the  ground  state energy  we
need to take into consideration  the  hard modes contribution.
 We should integrate  over hard
modes  using loop expansion  and find  a low  energy  effective
Lagrangian.
Here we consider the leading order of the expansion over number of
projections from the subspace $\{ E_\mu\}$ into the subspace
$\{ e_n\}$ and back. It was shown in Section 3 that an uncertainty of
this approximation is about a few percents at $\omega_0=\omega^*$  and
we  expect that an accuracy of our calculations will be about
few percents in this leading approximation.
{}From (\ref{b2}) and (\ref{b3}) we see that  in this approximation
there is no linear
terms for hard modes in the action.
It is easy to find one-loop effective potential for soft modes:
\bq
V^{(1)}(x_s)=\frac{1}{2\pi}\left(
\sqrt{12\lambda x_s^2}(\pi-2\arctan (\frac{\omega_0}{\sqrt{12\lambda
x_s^2}}))-\omega_0\ln (1+\frac{12\lambda x_s^2}{\omega_0^2})
\right)
\label{ef1}
\eq

Let us calculate the soft modes contribution into the energy
using 1-loop effective potential (\ref{ef1})
\be
\varepsilon^{1-loop}(\omega_0)=
\frac{\omega_0}{\pi}\left(1-\ln (
\omega_0\sqrt{\frac{2}{\pi}}
\int_0^\infty dB e^{-\frac{\pi}{\omega_0}V(B\sqrt{\frac{\omega_0}{\pi}})}
)\right)
\label{1loop}
\ee
here $V(x)=V^{1-loop}(x) =\lambda x^4+V^{(1)}(x)$.

The dependence of $\varepsilon^{1-loop}$ on $\omega_0$
is depicted in Fig.1 (line $(b)$). From Fig.1 we see that 1-loop hard mode
contribution is comparable with $\varepsilon_s$. Line $(c)$ in Fig.1
shows the dependence of $\varepsilon^{1-loop}(\omega_0)+
\varepsilon^{k1}_s(\omega_0)$.
Where $\varepsilon^{k1}_s$ is the leading kinetic term contribution.
And line $(d)$ in Fig.1 corresponds to
$\varepsilon^{2-loop}(\omega_0)$
which is calculated according eq.(\ref{1loop})
with 2-loop effective potential $V(x)$ for soft modes where
$V(x)=V^{2-loop}(x)=V^{1-loop}(x)+V^{(2)}(x)$ and the leading
kinetic term contribution is taken into account. The potential
$V^{(2)}(x)$ has the following form:
\bq
V^{(2)}(x)=\frac{1}{4\pi^2  x^2}
\left(\frac{\pi}{2}-\arctan (\frac{\omega_0}{\sqrt{12\lambda x^2}})
\right)^2
\label{v2}
\eq
\bq
-48\lambda^2x^2\int\frac{d\omega_1d\omega_2d\omega_3\delta (\omega_1+
\omega_2+\omega_3)}
{(2\pi)^2(\omega_1^2+12\lambda x^2)(\omega_2^2+12\lambda
x^2)(\omega_3^2+12\lambda x^2)}
\nonumber
\eq
where $\omega_1^2>\omega_0^2$, $\omega_2^2>\omega_0^2$,
$\omega_3^2>\omega_0^2$.

 In Fig.1 (line $(d)$) we see a very
weak dependence
of $\varepsilon^{2-loop} (\omega_0)$ on parameter $\omega_0$
in a large region
($\lambda^{1/3}<\omega^0<2.5\lambda^{1/3}$).
In this region the value of
the next to the leading corrections
is an order of variation $varepsilon (\omega_0)$.
and $\varepsilon^{2-loop} \simeq
(6.8\pm 0.3)\lambda^{1/3}$ which is in a good agreement
with exact result: $\varepsilon=0.66...$ (curve $(e)$ in Fig.1).
Thus we see a selfconsistence of the expansion in question.

\section{Discussion}

Let us discuss the main features of the approach.
 Two main assumptions are used here. The first one is that we
suppose that an expansion over the numbers of projections from
$\{ E_\nu\}$ to $\{ e_n\}$ and back does not diverge.
It was shown that the first correction of the expansion is
rather small in the case of the potential $\lambda x^4$ but
the general structure of this expansion is not known.
The second assumption is that there is a region for
the parameter $\omega_0$ where a perturbative expansion for
hard modes and a strong coupling expansion for soft mode
work at a same time.
The results obtained have shown a correctness of these assumptions
in the case of the potential $\lambda x^4$.
However, it is clear that this method does not work for a
potential which does not tend to infinity at $x\rightarrow\pm\infty$.
Also it is not possible to use this method (at least directly)
in instanton case due to the large kinetic term corrections
in soft modes sector. However this method gives a reasonable
results in the case of the potential considered here.

The next important problem is a question on translation invariance.
It is clear that this invariance is broken when we use basis
$E_\nu$. However in Section 1 it was shown that the path
integral in this basis is equal to the path integral in the
basis $e_n$ which does not break translational invariance.
The expansion over a numbers of projections from subspace
$\{ E_\nu\}$ into $\{ e_n\}$ and back corresponds to subtraction of
translational noninvariant contributions. In the case when
this expansion works we can control these contributions.

Here we considered the ground state energy only.
This parameter  is not convenient to study a restoration
of translational invariance.
In this context
it is interesting to investigate a propagator $<x(t_1)x(t_2)>$.
This question is very important for understanding
of applicability of the procedure.
The propagator has the following form
\bq
S(t_1,t_2)=\ll x(t_1),x(t_2)\gg
\label{p1}
\eq
\bq
=\ll B_\mu B_\nu \gg (E_\mu (t_1)-<E_\mu e_m^*>e_m(t_1))
(E_\nu (t_2)-<E_\nu e_n^*>e_n(t_2))
\nonumber
\eq
\bq
+\ll C_mC_n\gg e_m(t_1)e_n(t_2)
\nonumber
\eq
Here we take into account the shift $C_n\rightarrow C_n-B_\nu
<E_\nu e_n>$ which was introduced in the first Section,
$\ll \gg$ denotes the average value for the path integral
(see (\ref{b2},\ref{b3})). In eq.(\ref{p1}) we use that
in the leading order of the expansion over numbers of
projections
$\ll B_\mu C_n\gg =0$. It is obvious that $\ll C_n C_m\gg\sim
\delta_{nm}$ and the second term in eq.(\ref{p1}) depends on
$(t_1-t_2)$ only and does not break translational
invariance. In the leading order we have
 that $\ll B_\mu B_\nu\gg \sim \delta_{\mu\nu}$.
Then using (\ref{d4}) and (\ref{d5}) we obtain  that
the first term in (\ref{p1}) depends on $(t_1-t_2)$ only also.
It can be shown that  next to the  leading  corrections of  the
expansion do not break  the translational invariance.
Probably, that this procedure does not break
the invariance in any order of the expansion.

The most interesting application of the method is quantum field
theory. In this case a renormalization should be taken into consideration
by a standard way in an effective Lagrangian. For a gauge theory
it is necessary to study a question on a gauge invariance.

{\bf Acknowledgments}

Author thanks Max-Plank-Institute in Munich and Theory Group
of CEBAF for warm hospitality and useful discussions. I am
very grateful to participants of seminar in ITEP where
the main aspects of this paper were discussed.


\begin{thebibliography} {99}
\bibitem[*]{AA}Permanent address: ITEP, 117259 Moscow, Russia.
\bibitem{fei}R.P. Feynman, A.R. Hibbs, {\it Quantum mechanics and
Path Integrals} McGraw-Hill, New York (1965).
\bibitem{inst}A.A. Belavin, A.M. Polyakov, A.S. Shvarts, Yu.S. Tyupkin,
Phys.Lett. 59B (1975) 85; A.A. Belavin, A.M. Polyakov, Nucl.Phys.
B123 (1977) 429.
\bibitem{lat}R. Balian, J.M. Drouffe, C. Itzykson, Phys.Rev. D11 (1975)
2104; J. Banks, L. Susskind, J. Kogut, Phys.Rev. D13 (1976) 1043.
\bibitem{var}D.I. Diakonov, V.Yu. Petrov, Nucl.Phys. B245 (1984) 259.
\bibitem{susy}A.I. Vainshtein, V.I. Zakharov,  M.A. Shifman,
JETP Lett. 42 (1985) 224.
\bibitem{eff}S. Coleman, S. Weinberg, Phys.Rev. D7 (1973) 1888.

\end{thebibliography}
 \end{document}